\documentclass[aps,prl,twocolumn]{revtex4}

\usepackage{bm}
\usepackage{color}
\usepackage{graphicx}

\begin{document}

\title{Efficient entanglement distillation without quantum memory}

\author{Daniela Abdelkhalek $^{1,2}$, Mareike Syllwasschy $^{2}$, Nicolas J. Cerf $^{3}$, Jarom\'{i}r Fiur\'{a}\v{s}ek $^{4}$ and Roman Schnabel $^{1,2,\ast}$}

\affiliation{
$^{1}$ Institut f\"ur Laserphysik, Universit\"at Hamburg, Luruper Chaussee 149, 22761 Hamburg, Germany\\
$^{2}$ Institut f\"ur Gravitationsphysik, Leibniz Universit\"at Hannover and Max-Planck-Institut f\"ur Gravitationsphysik (Albert-Einstein-Institut), Callinstr.~38, 30167~Hannover, Germany\\
$^{3}$ Quantum Information and Communication, Ecole Polytechnique de Bruxelles, CP 165, Universit\'{e} libre de Bruxelles, 1050 Brussels, Belgium\\
$^{4}$ Department of Optics, Palack\'y University, 17. listopadu 12, 77146 Olomouc, Czech Republic\\
$^{\ast}$ Corresponding author: roman.schnabel@physnet.uni-hamburg.de}

\maketitle

\section{Abstract}
Entanglement distribution between distant parties is an essential component to most quantum communication protocols. Unfortunately, decoherence effects such as phase
noise in optical fibers are known to demolish entanglement. Iterative (multistep) entanglement distillation protocols have long been proposed to overcome decoherence,
but their probabilistic nature makes them inefficient since the success probability decays exponentially with the number of steps. Quantum memories
have been contemplated to make entanglement distillation practical, but suitable quantum memories are not realised to date. Here, we present the theory for an
efficient iterative entanglement distillation protocol without quantum memories and provide a proof-of-principle experimental demonstration. The scheme is applied to
phase-diffused two-mode-squeezed states and proven to distill entanglement for up to three iteration steps. The data are indistinguishable from those an efficient scheme
using quantum memories would produce. Since our protocol includes the final measurement it is particularly promising for enhancing continuous-variable
quantum key distribution.

\section{Introduction}

Light is the most suitable carrier of quantum information over long distances. Distribution of entangled states of light among distant nodes of a quantum communication network \cite{Kimble2008,Weedbrook2012} may be used for various purposes, such as quantum cryptography \cite{Ekert91,Scarani2009} or quantum teleportation \cite{Bouwmeester1997,Furusawa1998}. However, the transmission of quantum states of light is, in practice, unavoidably affected by losses and other decoherence effects that usually grow with distance. To fight these detrimental effects, entanglement distillation can be employed, which extracts from a large number of noisy and weakly entangled states a smaller number of copies with increased entanglement and purity 
\cite{Bennett1995,Deutsch1996,Kwiat2001,Pan2003,Hage2008,Dong2008,Takahashi2010,Hage2010,Kurochkin2014}. Crucially, entanglement distillation only requires local quantum operations and classical communication between the spatially separated parties holding parts of the entangled state. 
\begin{figure}[b]
\includegraphics[width=\linewidth]{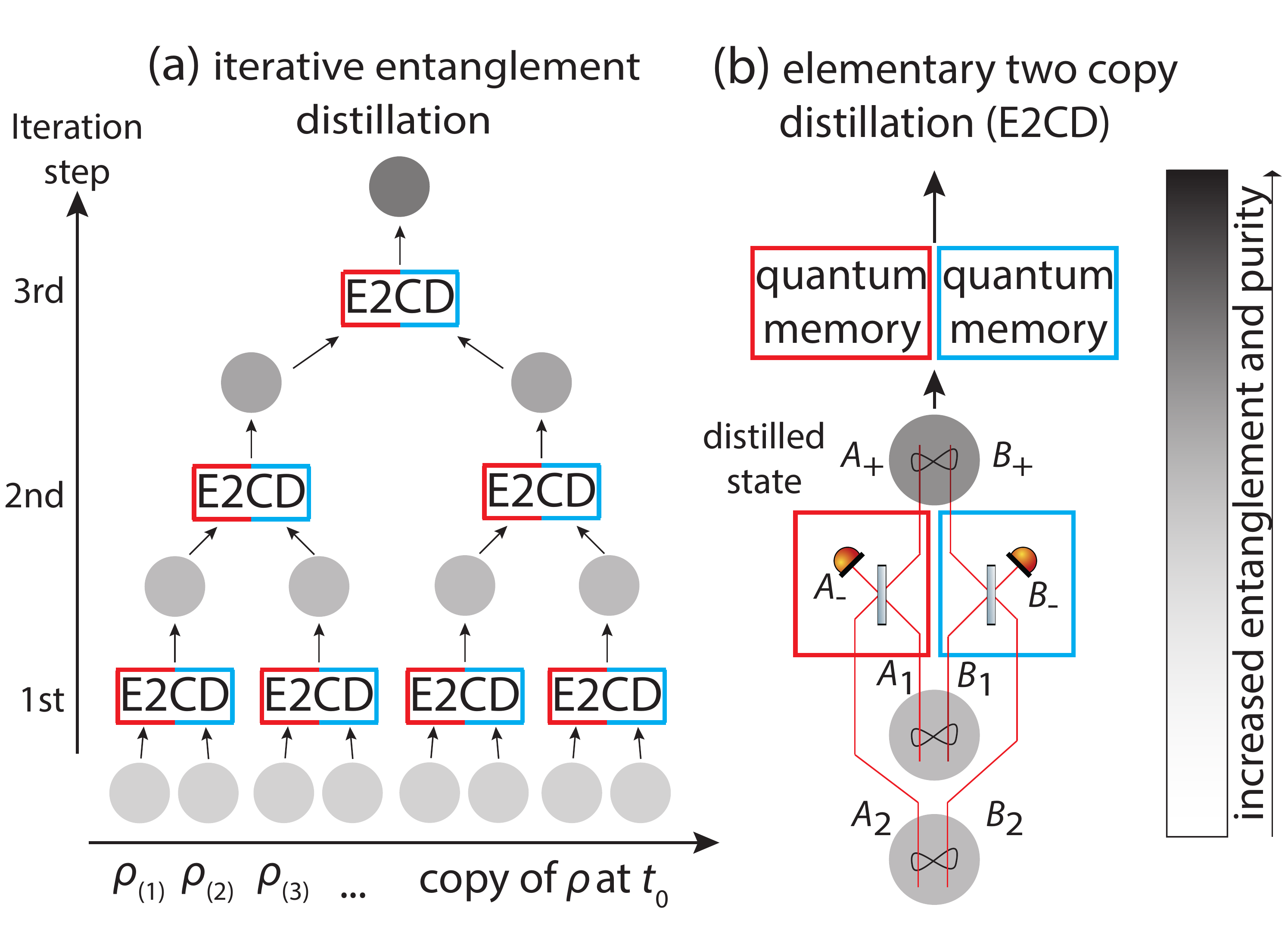}
\caption{\textbf{Iterative entanglement distillation.} \textbf{(a)} At each elementary step of the protocol, 
two copies of the input state, described by the density matrices $\rho$, are combined to produce, with some probability, one output copy with better properties.
\textbf{(b)} Elementary step of iterative Gaussification of two-mode continuous variable quantum states.
Alice's and Bob's modes locally interfere on balanced beam splitters and one output mode on each side is measured with a photodetector. 
The procedure succeeds if both modes are projected onto vacuum. A quantum memory was previously deemed necessary for
an efficient implementation of such an iterative protocol as one must store the successfully distilled states in a quantum memory to make them available for subsequent distillation steps.}
\label{fig:schematic}
\end{figure}
A canonical iterative entanglement distillation protocol \cite{Bennett1995,Deutsch1996} is illustrated in Fig. \ref{fig:schematic}(a). At each step, two copies of a decohered entangled state are consumed and, 
with some probability, one copy of a distilled state with improved properties is produced, which in turn serves as the input for the next round of the protocol. 
An efficient implementation of this iterative protocol requires a quantum memory \cite{Simon2010}, which motivates the extensive current effort for harnessing the light-to-matter interface, 
ultimately leading to quantum repeaters \cite{Duan2001}. Without quantum memory, all elementary two-copy distillation steps need to succeed simultaneously, 
which imposes an exponential overhead in terms of required resources and hence drastically reduces the success rate of the protocol. 
Let us consider $N$ iterations of the protocol and suppose, for simplicity, that the success probability of each elementary distillation step is the same and equal to $P$. A single attempt to perform $N$ iterations requires $2^N$ input states, and all $2^{N}-1$ elementary distillation steps must succeed simultaneously. To obtain $1$ distilled copy, one thus needs to consume on average $2^N/P^{2^{N}-1}$ input states. With quantum memories, the distillation becomes far more efficient because the successfully distilled states after each step of the protocol can be stored and employed in the subsequent step as required. 
Ideally, this reduces the required number of input states to $2^N/P^N$. Note that the hardware-efficient pumping Gaussifier proposed in \cite{Campbell2013} requires only a single quantum memory unit on each side and enables sequential processing of the individual copies of distilled quantum state but achieves similarly inefficient distillation rate as a purely optical scheme without quantum memory.

To date, single-copy entanglement concentration \cite{Kwiat2001,Dong2008,Takahashi2010,Kurochkin2014,Ulanov2015} and elementary two-copy entanglement distillation \cite{Pan2003,Hage2008} have been demonstrated for both discrete- and continuous-variable quantum states of light.  A collective three-copy distillation of continuous-variable entangled states could even be implemented \cite{Hage2010}, but it remained inefficient in the absence of a quantum memory. Aside from this, an efficient realisation of the full iterative multicopy entanglement distillation protocol could never been done because it is pending on operating an efficient quantum memory. 

Here, we address this challenge from an opposite viewpoint and cancel out the need for quantum memories by exploiting the recently proposed concept of emulation of a quantum protocol \cite{Fiurasek2012,Walk2013}. The emulation replaces the actual physical implementation of a certain quantum operation by suitable postprocessing of measurement data and postselection, 
which offers a high potential to circumvent hardware implementation problems. In particular, the emulation of noiseless quantum amplification and attenuation by processing the data resulting from (eight-port) homodyne detection was proposed \cite{Fiurasek2012,Walk2013}, and a proof-of-principle experimental emulation of single-mode noiseless quantum amplification was reported \cite{Chrzanowski2014}.

In this article, we apply such a strategy to multi-copy continuous-variable entanglement distillation. 
Our scheme is solely based on measurement data taken on a single beam of light that carries a continuous stream of copies of a decohered entangled state. Remarkably, due to the specific nature of the measurement and post-processing, neither quantum memories nor the simultaneous physical realisation of many copies are required, in contrast with common knowledge on entanglement distillation. In principle, an arbitrary number of elementary (2-copy) distillation steps can be emulated by post-processing the experimental data, solely limited by the length of the measured data stream.
Specifically, we demonstrate the distillation of phase-diffused two-mode squeezed states of light by iterative Gaussification \cite{Browne2003,Eisert2004,Datta2012,Campbell2012} based on interference 
of two copies of the state on balanced beam splitters, followed by projecting one output port on each side onto a vacuum state, 
see Fig. \ref{fig:schematic}(b). We show that this scheme can be emulated by suitable processing of data obtained by eight-port homodyne detection on each mode of each phase-diffused state.
Crucially, such emulation is completely indistinguishable from a full physical implementation with even ideal quantum memories to anyone outside Alice and Bob's labs. The only difference is that the distilled states are already detected and not physically available for further processing.

Nevertheless, the data may be used to fully characterise the distillation protocol and, e.g., to extract a secret key from the distilled states. Our procedure is therefore particularly suitable for quantum cryptography, where it can convert seemingly useless highly noisy states into states that allow for the extraction of a secret key. It is similarly applicable to all related quantum communication protocols provided they terminate with suitable measurements.

\begin{figure} [t]
\includegraphics[width=\linewidth]{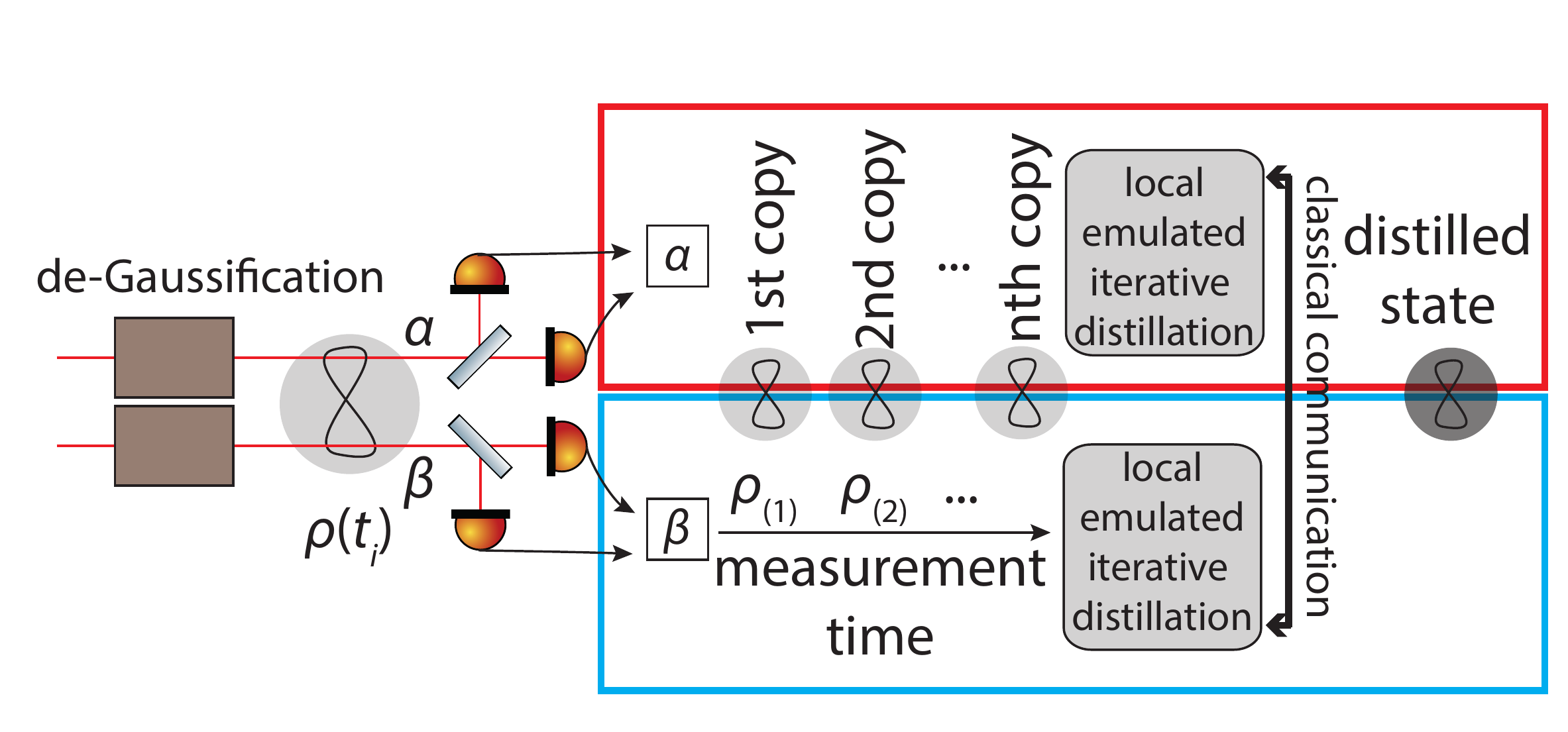}
\caption{\textbf{Emulation of iterative Gaussification protocol.} A single, continuously operated source distributes copies of the entangled and decohered states $\rho_0$ to Alice and Bob. The states' coherent amplitudes are locally measured with two out-of-phase balanced homodyne detectors each, so called 'eight-port homodyne detectors'. By prolonging the measurement time any quantity of copies may be generated. An emulated iterative distillation protocol based on local operations and classical communication is realized as described in the main text. In the end Alice and Bob share data that is equivalent to that from measurements on distilled entangled states.}
\label{fig:newprotocol}
\end{figure}

\section{Results}

\textbf{Emulation of iterative entanglement distillation.} The starting point of our work is the iterative Gaussification scheme \cite{Browne2003,Eisert2004} illustrated in Fig.~\ref{fig:schematic}, which can serve for entanglement distillation of non-Gaussian quantum states of light. Each elementary step of the Gaussification protocol requires two copies of a two-mode entangled state, 
whose modes are labeled as $A_1$, $B_1$, and $A_2$, $B_2$, respectively. Modes $A_1$ and $A_2$ belong to Alice, and modes $B_1$ and $B_2$ are held by Bob. 
Both Alice and Bob combine their pairs of modes on a balanced beam splitter and obtain output modes $A_{+}$, $A_{-}$ and $B_{+}$, $B_{-}$, 
where $+$ and $-$ refers to constructive and destructive interference, respectively. Alice and Bob perform photon number measurement 
on the modes $A_{-}$ and $B_{-}$ and exchange the measurement outcomes. This elementary distillation step is successful 
if both modes $A_{-}$ and $B_{-}$ are projected onto a vacuum state, and the distilled state in modes $A_{+}$ and $B_{+}$ provides the input for another round of the iterative protocol. 
Suppose now that Alice and Bob decide to measure the final distilled state with eight-port homodyne detectors, which perform projections onto coherent states. The generally impracticable implementation of the full iterative distillation protocol can be replaced by the following emulation procedure as depicted in Fig.~\ref{fig:newprotocol}. 
Alice and Bob perform the eight-port homodyne detection directly on the distributed decohered entangled state, and they repeat this measurement many times to collect a sufficiently large data set. Let $\alpha_j$ and $\beta_{j}$ denote the measurement outcomes (complex amplitudes of coherent states)
for the $j$-th copy of the distributed state. Alice and Bob then combine the measurement results into pairs, say $(\alpha_{2n},\alpha_{2n+1})$ and $(\beta_{2n},\beta_{2n+1})$, and emulate the interference on beam splitters making the first layer of the protocol by performing simple additions and subtractions of the measurement results, 
which faithfully mimics interference of coherent states,
\begin{equation}
 \begin{array}{lcl}
 \alpha_{n,+}=\frac{1}{\sqrt{2}}(\alpha_{2n}+\alpha_{2n+1}), & \quad & \alpha_{n,-}=\frac{1}{\sqrt{2}}(\alpha_{2n}-\alpha_{2n+1}), \\[2mm]
 \beta_{n,+}=\frac{1}{\sqrt{2}}(\beta_{2n}+\beta_{2n+1}), & \quad & \beta_{n,-}=\frac{1}{\sqrt{2}}(\beta_{2n}-\beta_{2n+1}).
 \end{array}
\end{equation}

Projection of the output modes $A_{-}$ and $B_{-}$ onto vacuum is equivalent to conditioning on $\alpha_{n,-}=\beta_{n,-}=0$.
After the conditioning we obtain a reduced set of pairs $\alpha_j^{(1)}=\alpha_{n_j,+}$ and $\beta_j^{(1)}= \beta_{n_j,+}$, where $n_j$ represent the values of $n$ for which the conditioning was successful. These pairs of data represent effective measurement outcomes of eight-port homodyne detections on modes A and B of the distilled two-mode state 
after the first iteration of the distillation protocol. This procedure can be repeated, and outputs of the $k$-th round of the protocol, $\alpha_j^{(k)}$ and $\beta_j^{(k)}$, can be used as inputs of the next round of the protocol, resulting in $\alpha_j^{(k+1)}$ and $\beta_j^{(k+1)}$.

{Since the probability to obtain the specific outcomes $\alpha_{n,-}=\beta_{n,-}=0$ in eight-port homodyning vanishes, 
we need to modify the conditioning to make the protocol practicable. A theoretically appealing modification is to impose an acceptance probability that is a Gaussian function of the complex amplitudes,
\begin{equation}
P_{\mathrm{acc}}(\alpha_{-},\beta_{-})= \exp\left(-\frac{|\alpha_{-}|^2}{\bar{n}}\right)\exp\left(-\frac{|\beta_{-}|^2}{\bar{n}}\right).
\end{equation}
Physically, this corresponds to a modified conditioning in Fig. 1(b), where modes $A_{-}$ and $B_{-}$ are projected onto thermal states 
with a mean number of thermal photons equal to $\bar{n}$. The choice of Gaussian $P_{\mathrm{acc}}$ guarantees that the iterative Gaussification protocol indeed converges to a Gaussian state \cite{Eisert2004,Campbell2012}, which greatly simplifies its theoretical treatment and allows us to derive analytically the asymptotic state to which the protocol converges.

To be more specific, let us consider as an example the distillation of phase-diffused two-mode squeezed states, as employed in our experiment.
A covariance matrix of the initial Gaussian symmetric two-mode squeezed state before phase diffusion reads \cite{Weedbrook2012}
\begin{equation}
\gamma_{\mathrm{AB}}= 
\left(
\begin{array}{cccc}
a & 0 & b & 0 \\
0 & a & 0 & -b \\
b & 0 & a & 0 \\
0 & -b & 0 & a
\end{array}
\right).
\label{gmAB}
\end{equation}
Here $a$ denotes the (symmetric) variance of phase space projections of the complex amplitude (quadratures) of the individual modes, and $b$ represents the correlations between quadratures of modes $A$ and $B$. 
Suppose now that modes A and B are sent through noisy channels where random phase shifts $\phi_{\mathrm{A}}$ and $\phi_{\mathrm{B}}$ are imposed. For the sake of simplicity, we shall assume that the phase diffusions are independent and have the same statistics for both modes (though our results can be easily extended to more general scenarios). Note that the phase noise does not modify the value of parameter $a$ since each mode is locally in a thermal state that is invariant under phase shift. So the covariance matrix of the phase diffused non-Gaussian state at the output of the noisy channels preserves the form (\ref{gmAB}), but the phase diffusion reduces the intermodal correlations,
\begin{equation}
a_{\mathrm{PD}}=a, \qquad b_{\mathrm{PD}}=qb,
\end{equation}
where $q=\langle \cos\phi_{\mathrm{A}} \cos\phi_{\mathrm{B}}\rangle$ quantifies the phase diffusion for uncorrelated noise. The stronger the phase diffusion, the smaller is $|q|$. We assume here 
that the phase noise is symmetric, $\langle \sin\phi_\mathrm{A}\rangle=\langle \sin\phi_\mathrm{B}\rangle=0.$
If no phase diffusion is present ($\phi_\mathrm{A}$ and $\phi_\mathrm{B}$ always zero) $q$ equals unity. If the phases are fully random $q$ equals zero.
For given fixed values of the random phase shifts $\phi_\mathrm{A}$ and $\phi_\mathrm{B}$, the covariance matrix $\gamma_{\mathrm{AB}}$ of an input two-mode state is transformed according to $R\gamma_{\mathrm{AB}} R^T$ with
\begin{equation}
R(\bm{\phi})=
\left(
\begin{array}{cccc}
\cos\phi_\mathrm{A} & \sin\phi_\mathrm{A} & 0 & 0 \\
-\sin\phi_\mathrm{A} & \cos\phi_\mathrm{A} & 0 & 0 \\
0 & 0 & \cos\phi_\mathrm{B} & \sin\phi_\mathrm{B} \\
0 & 0 & -\sin\phi_\mathrm{B} & \cos\phi_\mathrm{B}
\end{array}
\right)
\end{equation}
describing the action of the particular phase shift values $\phi_\mathrm{A}$ and $\phi_\mathrm{B}$ at the level of the covariance matrix. 
This expression has to be averaged over the random phase shifts to obtain the resulting covariance matrix of the de-phased state. Note that in this last averaging step we assume that the mean values of quadratures of the input state vanish, which is satisfied in our experiment where we utilize squeezed vacuum state. If non-zero, the mean values can be eliminated and set to zero by suitable local coherent displacements, which in our emulation-based scheme can be implemented simply as displacements of the measured data.
With the help of the general theory of iterative Gaussification protocols presented in Ref. \cite{Campbell2012}, one can derive the following 
analytical expression for the covariance matrix of the asymptotic Gaussian state,
\begin{equation}
\gamma_{\mathrm{AB},\infty}=\left\langle R(\bm{\phi})[\gamma_{\mathrm{AB}}+(2\bar{n}+1)I]^{-1} R^{T}(\bm{\phi})\right\rangle_{\phi}^{-1}-(2\bar{n}+1)I.
\label{gmDist}
\end{equation}
Here $I$ denotes the identity matrix and $\langle .\rangle_{\phi}$ represents statistical averaging over random phase diffusions.

\begin{figure}[t]
\includegraphics[width=\linewidth]{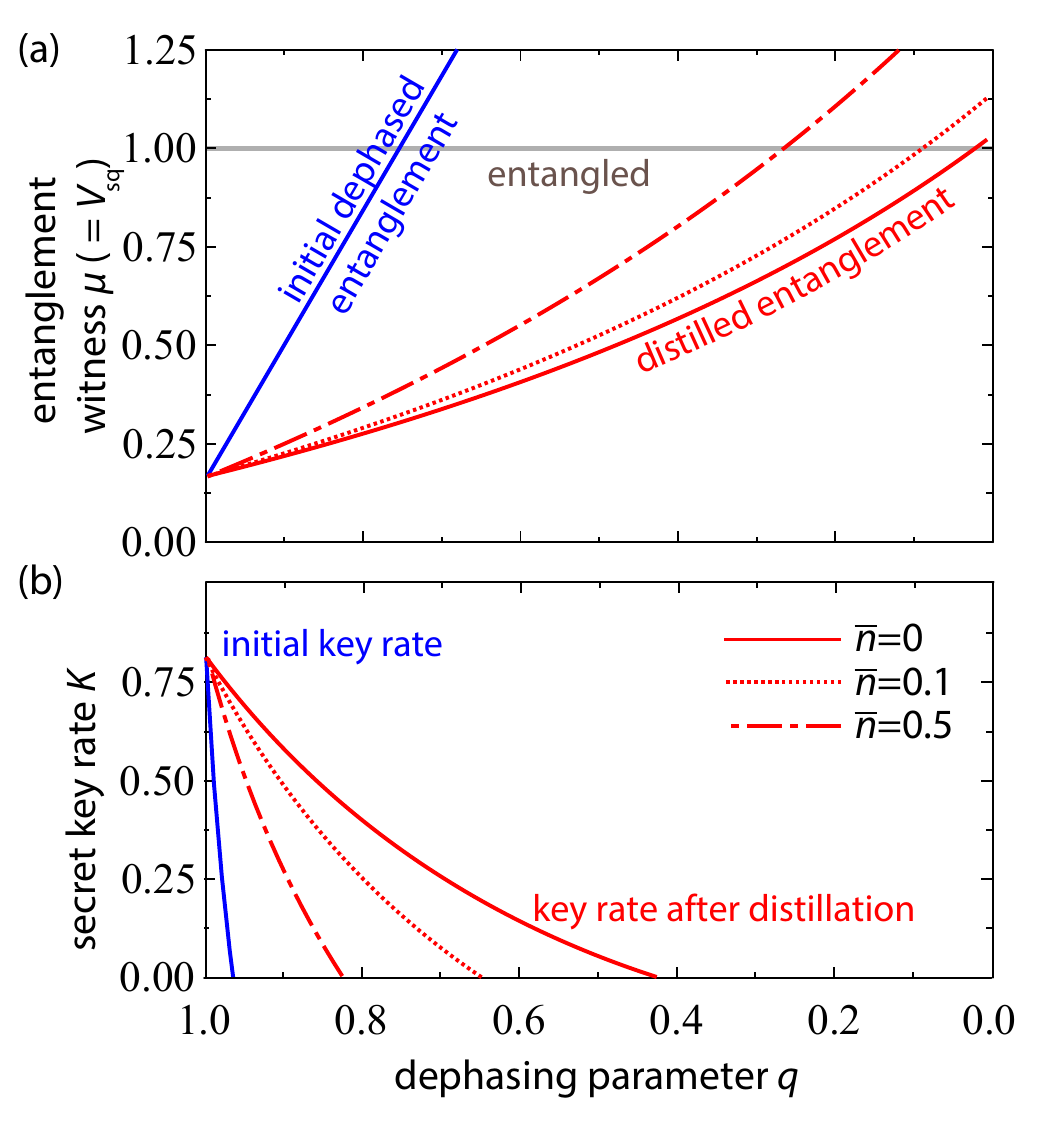}
\caption{\textbf{Entanglement distillation of phase-diffused two-mode squeezed states.} Squeezing variance $V_{\mathrm{sq}}$ (a) and distillable 
secret key $K$ (b) are plotted for the phase-diffused state (blue) and for the asymptotic distilled state for different $\bar{n}$ (red lines) corresponding to Eq.(\ref{gmDist}). $\bar{n}$ gives here the mean number of thermal photons of the state we project onto.
The parameter $q$ quantifies the strength of the phase diffusion: $q=1$ indicates that no phase noise is present whereas $q=0$ corresponds to a completely randomized phase. The parameters of the input state before phase diffusion
read $a=3.583$ and $b=3.417$, which corresponds to variances of squeezed and anti-squeezed quadratures $V_{\mathrm{sq}}=1/6$ and $V_{\mathrm{antisq}}=7$. 
}
\label{keyrate}
\end{figure}

The performance of the distillation protocol and its usefulness for quantum key distribution is illustrated in Fig.~\ref{keyrate}. The squeezing properties of the two-mode state can be characterized by the squeezed variance $V_{\mathrm{sq}}$ defined as the minimum eigenvalue of the covariance matrix,
$V_{\mathrm{sq}}=\min(\mathrm{eig}(\gamma_{\mathrm{AB}}))$, while entanglement can be detected using the Duan-Simon criterion \cite{Duan2000,Simon00}.
Technically, one has to determine the minimum symplectic eigenvalue $\mu$ of a covariance matrix corresponding to the partially transposed state, and entanglement is witnessed if $\mu<1$. For a symmetric two-mode squeezed state as in Eq. (\ref{gmAB}), we find that $V_{\mathrm{sq}}=\mu=a-|b|$, hence the presence of squeezing, 
$V_{\mathrm{sq}}<1$, also indicates that the state is entangled. In Fig.~\ref{keyrate}(a), the dependence of $\mu$ on the dephasing parameter $q$ is plotted for the phase-diffused state (blue solid line) and asymptotic distilled state for different $\bar{n}$ (red lines). The enhancement of entanglement by distillation is clearly visible, and the closer we choose $\bar{n}$ to zero (which corresponds to projection on vacuum) the stronger is the effect. 
Importantly, the distillation protocol can even recover seemingly lost quadrature entanglement as the distilled state can exhibit $\mu<1$ even if the initial phase-diffused state exhibited $\mu>1$. Distillation of entanglement necessarily implies that the input state was (weakly) entangled too. However, its initial entanglement is not visible via the covariance matrix (certifying a separable Gaussian state) but hidden in higher-order correlations of the quadratures.

We can determine the maximum tolerable phase noise (minimum $|q|$) for which we can still distill Gaussian entanglement. 
For the \emph{pure} symmetric two-mode Gaussian state with covariance matrix given by Eq. (\ref{gmAB}), and  $a=\cosh 2r$ and $b=\sinh 2r$ 
we find that the Gaussian entanglement is asymptotically distilled provided that
\begin{equation}
\frac{\bar{n}}{\bar{n}+1}\tanh r< |q| .
\end{equation}
Here $r$ denotes the squeezing parameter of the squeezed input states. If we project on vacuum ($\bar{n}=0$) then entanglement is distilled for any $r>0$ and $|q|>0$. For $\bar{n}>0$, however, entanglement is only distilled in the asymptotic limit if the dephasing is not too large ($|q|$ not too small). 
For mixed states, the entanglement might not be distillable even if $\bar{n}=0$, c.f. the red solid line in Fig. \ref{keyrate}(a) which crosses the horizontal line $\mu=1$ at nonzero $q$.

We can go further and calculate the distillable secret key in continuous-variable quantum key distribution $K=I_{\mathrm{AB}}-\chi_{\mathrm{AE}}$ (Fig.~\ref{keyrate}(b)), where $I_{\mathrm{AB}}$ is the mutual information between Alice's and Bob's data and $\chi_{\mathrm{AE}}$ is Eve's classically accessible information on Alice's quantum state. The key rate can be calculated analytically using the optimality of Gaussian attacks \cite{Grosshans2004,Navascues2006,Garcia2006} and the well known formulas for entropies of Gaussian probability distributions and Gaussian quantum states (see e.g. \cite{Weedbrook2012,Fiurasek2012,Lodewyck2007} for more details).
 As shown in Fig.~\ref{keyrate}(b), the distillation protocol can significantly enlarge the range of $q$ values for which a secret key can be distilled. Crucially, the key rate for the distilled state can be positive even if the initial phase-diffused state does not exhibit squeezing, $V_{\mathrm{sq}}>1$, that is, if its covariance matrix is compatible with that of a Gaussian separable state. In such a case, no secure key could be obtained from the decohered state by any protocol involving Gaussian measurements and a security analysis based on the covariance matrix of that state. The emulation of iterative entanglement distillation can thus efficiently convert seemingly useless noisy data into data from which a non-zero secret key could be extracted.\\
 \\
\begin{figure} [b]
\includegraphics[width=\linewidth]{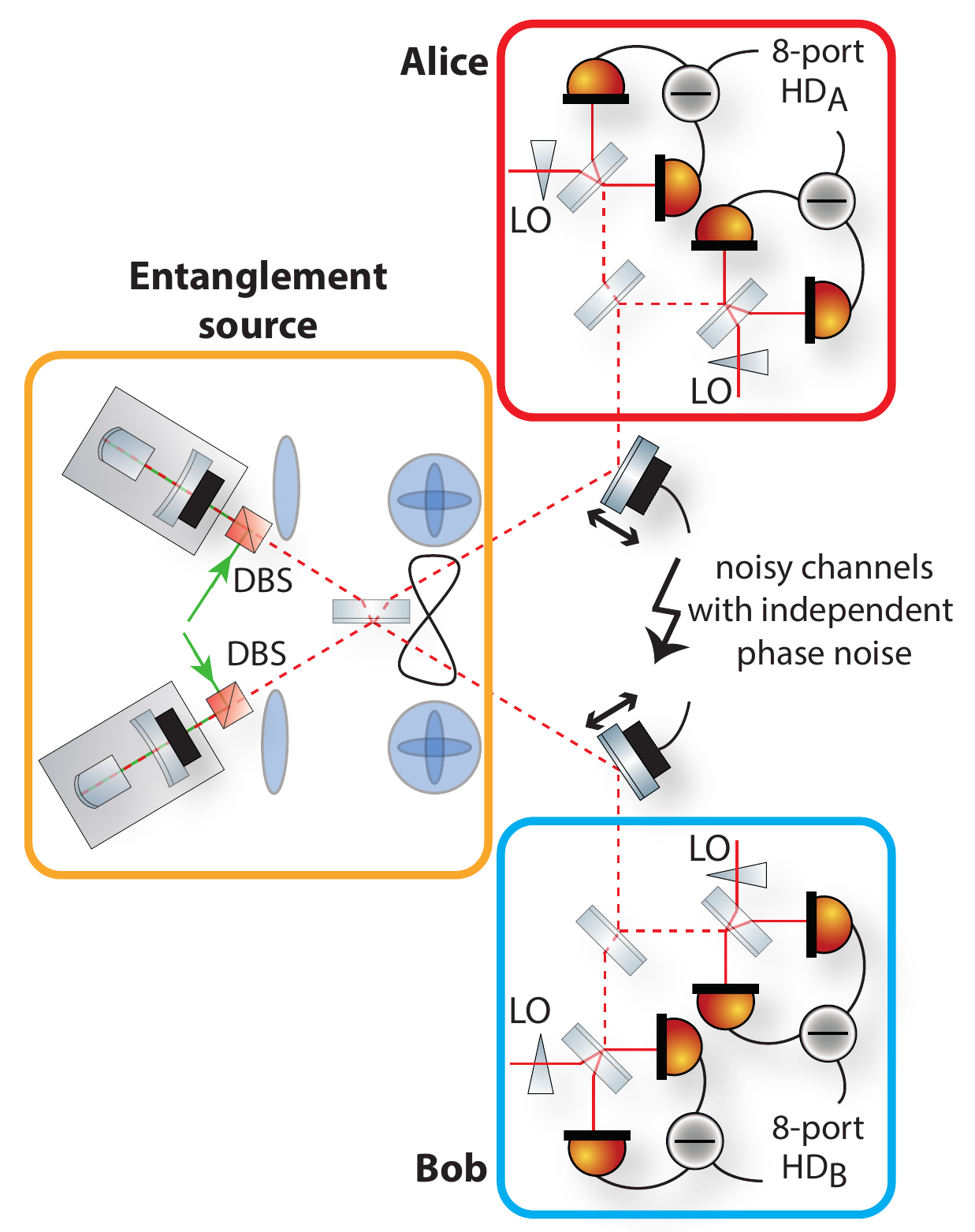}
\caption{\textbf{Experimental setup.} Two propagating squeezed light modes are superimposed on a balanced beam splitter. The entangled outputs are sent to two separate sites. During transmission the modes are exposed to phase noise which produces de-Gaussified mixed modes. For strong phase noise the two-mode squeezing, i.e. the entanglement in the second moments of the quadratures, is completely destroyed. The decohered modes are continuously and unconditionally detected by eight-port homodyne detectors, respectively, which have negligible intrinsic phase noise and detection efficiencies of about 87$\pm$6$\%$. DBS: dichroic beam splitter.}
\label{fig:setup}
\end{figure}

\textbf{Experimental setup.} A schematic picture of the experimental setup is shown in Fig. \ref{fig:setup}. In a first step, we generated two gaussian entangled light fields, which was achieved by superimposing two squeezed vacuum states on a balanced beam splitter with a 90\,$^\circ$ phase-shift \cite{Furusawa1998,Eberle2013}. 
Previous realisations of multicopy continuous-variable entanglement distillation protocols employed so-called v-class entangled states \cite{Hage2008,Hage2010}, which are simply achieved by superimposing a squeezed vacuum mode and an ordinary vacuum mode on a balanced beam splitter. Replacing the vacuum mode by a second squeezed light field increases the experimental effort but also allows in principle stronger entanglement, which is needed for QKD. Our scheme has the advantage that the (efficient) multistep iterative distillation of entanglement only requires at most two squeezed light sources, whereas an actual hardware implementation of a (still inefficient) iterative distillation without quantum memories requires doubling the number of squeezed light sources with each iteration step. 
In our proof-of-principle experiment the squeezed-vacuum sources produced slightly different squeezing values, around 3\,dB. Due to this asymmetry the covariance matrix of the entangled state didn't hold the form given by Eq.(\ref{gmAB}) but was given by 
\begin{equation}
\gamma_{\mathrm{AB,exp}}= 
\left(
\begin{array}{cccc}
3.20 & -0.13 & -2.90 & -0.04 \\
-0.13 &  6.24 & -0.03 & 6.08 \\
-2.90 & -0.03 &  3.70 & -0.06 \\
-0.04 & 6.08 & -0.06 & 6.83 \\
\end{array}
\label{gamma_exp}
\right).
\end{equation}
We reconstructed this covariance matrix by measuring the prepared state without additional phase diffusion with the eight-port homodyne detectors and using the formula $\gamma_{\mathrm{AB,exp}}=2\gamma_{\mathrm{EHD}}-I$, where 
$\gamma_{\mathrm{EHD}}$ is a covariance matrix calculated directly from the eight-port homodyne data, and $I$ denotes the identity matrix.
The small values at the off-diagonal elements are not exactly zero due to imprecisions in the quadrature phases. 
The different values of the diagonal elements are mainly caused by the different squeezing values. The matrix $\gamma_{\mathrm{AB,exp}}$ was reconstructed without correcting for detection inefficiencies and detector dark noise.
 For this reason, the covariance matrix is a good description of the actual measurement data (but a less good description of the quantum state before detection.)
We note that two-mode squeezing of more than 10\,dB has already been demonstrated \cite{Eberle2013} and our scheme would work equally well with these higher squeezing strengths.

Our entangled modes were distributed between Alice and Bob's sites. During transmission, the modes were exposed to independent phase-noise modeling a noisy transmission through, e.g., optical fibers. The noise was applied by varying the position of steering mirrors in the path, driven by piezo electric transducers. We varied the strength of the noise by amplifying or attenuating the voltage given to the actuators to examine the behavior of the protocol for different noise strengths. For all measurements shown in the next section we used the same phase diffusion which corresponded to $q=0.78$. 
At the sites, the received light beams were absorbed in eight-port homodyne detectors, consisting of a balanced beam splitter and two conventional balanced homodyne detectors, which measure the amplitude and phase quadrature on the first and the second beam, respectively. This measurement corresponds to a projection onto coherent states. The recorded data provided the basis to execute the emulated distillation protocol.

\begin{figure} [h]
\includegraphics[width=\linewidth]{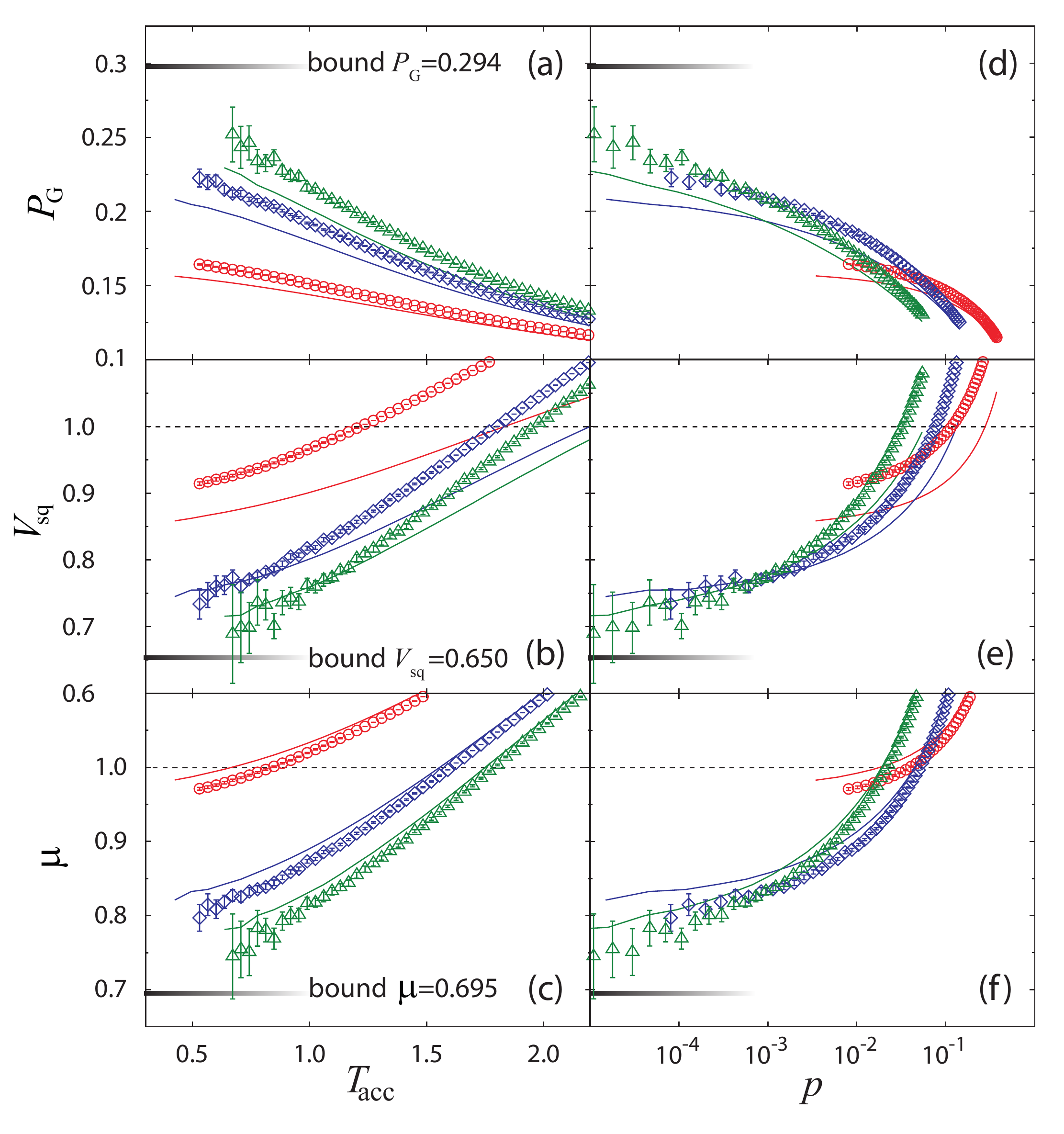}
\caption{\textbf{Experimental distillation results.} The panels show the dependence of $V_{\mathrm{sq}}$, $\mu$ and $P_\mathrm{G}$ on the threshold $T_{\mathrm{acc}}$ 
and success probability $p$ for one (red, circle), two (blue, square), and three (green, triangle) iterations of the distillation protocol as discussed in the main text. The error bars give a $2\sigma$ confidence interval and are obtained by bootstrapping the samples 100 times where the length of each bootstrapping sample equals the length of the analysed data set. The solid lines give theoretical predictions obtained by Monte Carlo simulations for $10^{10}$ samples. The simulation uses the undiffused measured covariance matrix (Eq.\ref{gamma_exp}) and the experimentally determined phase noise factor $q$=0.78. An uncorrelated Gaussian distribution of the random phase shifts with equal variances is assumed, and the model takes into account additional losses induced by our implementation of the de-phasing channel.	Since the simulated curves strongly depend on the exact statistics of the phase noise, the theoretical curves for $V_{\mathrm{sq}}$ show some discrepancies with the experimental data. By a slight change of $q$ we could get a good fit for $V_{\mathrm{sq}}$ at the expense of a worse fit for $\mu$. However the general good agreement suggests that our assumed model is a good approximation. The short fading horizontal lines represent the asymptotic limit for projection onto vacuum ($\bar{n}=0$).}
\label{fig:results}
\end{figure}
\textbf{Experimental results.} In the experiment, we collected up to $5\cdot 10^8$ data points for a fixed strength $q$ of phase noise, and we emulated up to three rounds of the iterative entanglement distillation protocol. In order to simplify data processing, we employed a deterministic conditioning rule and the elementary entanglement distillation step was taken successful if $|\alpha_{-}|< T_{acc}$ and $|\beta_{-}|< T_{acc}$, where $T_{\mathrm{acc}}$ is a tunable threshold. While such conditioning may lead to some residual non-Gaussianity of the asymptotic state, it does not modify the qualitative properties of the distillation protocol. We have used the resulting data to reconstruct the covariance matrix $\gamma$ of the two-mode state after each iteration of the distillation protocol, and we have used the covariance matrix to determine various properties of the distilled state. The results are plotted in Fig.~\ref{fig:results}, which shows the two-mode squeezing variance $V_{\mathrm{sq}}$, the symplectic eigenvalue $\mu$ witnessing entanglement, and the quantity $P_G=1/\sqrt{\det{\gamma}}$, which for Gaussian states coincides with the state purity. 

Panels \ref{fig:results} (a,b,c) illustrate that each iteration of the protocol increases squeezing and (Gaussian) entanglement of the distilled state  for a fixed threshold $T_{\mathrm{acc}}$, as indicated by the reduction of $V_{\mathrm{sq}}$ and $\mu$. Furthermore, $P_\mathrm{G}$ increases with the number of iterations, which is a strong indication that the purity of the state is increased by the protocol. 
 The absolute bounds presented in all panels are the asymptotic bounds for projection onto vacuum ($\bar{n}=0$) calculated with Eq. (\ref{gmDist}). Statistical properties of the random phase shifts required for evaluation of the phase average in Eq.~(\ref{gmDist}) were determined by comparing the covariance matrices of the initial and the de-phased states.
It is remarkable how close the data come to the absolute bound with only 3 iteration steps and a finite threshold $T_{\mathrm{acc}}$.
Panels \ref{fig:results} (d,e,f) show the practically more relevant dependence of $V_{\mathrm{sq}}$, $\mu$, and $P_\mathrm{G}$ on the total success probability of the protocol $p$, which was determined as the ratio of the number of distilled copies of the state versus the total number of input copies of the state. These plots fully demonstrate the usefulness of an iterative distillation scheme. For a fixed total success probability, more iterations may be advantageous as they lead to better squeezing, entanglement and purity. 
The crossover between one and two iterations is clearly and unambiguously demonstrated by the data, while the crossover between two and three iterations can be seen in the region of $10^{-4}<p<10^{-3}$, albeit the results in this region are already affected by statistical uncertainties. We have repeated the experiment for several different strengths of phase noise, and the results were in all cases qualitatively similar to those shown in Fig.~5.\\

We emphasize that the emulated iterative entanglement distillation presented here is inherently as efficient as if the protocol had been implemented with quantum memories, which would store successfully distilled states from previous rounds of the protocol to be utilized in the next rounds of the protocol. In the emulation, we naturally use in the subsequent rounds of the protocol only the successfully distilled states from previous rounds, which ensures that the average number of input copies that are consumed to produce a single distilled copy 
scales exactly as if quantum memories had been used.

\section{Discussion}

Our proof-of-principle experiment paves the way to efficient quantum communication protocols, where a specific data processing mimicks quantum memories (hence, provides the same advantage in terms of resources) without actually requiring them. This strategy can be applied at the end points of any entanglement distribution scheme where the end users perform eight-port homodyne detection. Importantly, our protocol is not limited to schemes where entanglement is physically distributed, but is also applicable to a prepare-and-measure quantum key distribution schemes where Alice prepares Gaussian modulated coherent states and sends them to Bob who performs eight-port homodyne detection \cite{Weedbrook2004}. Indeed, since the preparation of coherent states by Alice is indistinguishable from preparation of two-mode squeezed vacuum followed by eight-port homodyne detection on Alice's mode, the preparation can be equivalently interpreted as a measurement on a (virtual) shared entangled state. Similarly, our procedure is also applicable to the recently proposed and demonstrated measurement-device independent continuous-variable QKD protocol \cite{Guo2014,Pirandola2015}, where Alice and Bob both prepare randomly modulated coherent states and send them to an untrusted relay which performs a continuous-variable Bell measurement on the two modes and publicly announces the results of the measurement.
Given its wide range of potential applications, our proposal represents a promising tool for improving the performance of various quantum communication systems.

\section{Methods}

\textit{Squeezed light preparation:} The main light source was a Nd:YAG laser that produced a continuous-wave field at a wavelength of 1064\,nm, as well as frequency-doubled light at 532\,nm. The infrared light provided the control fields for the active length stabilisation of two squeezed light resonators containing 7\,\% magnesium-oxid-doped lithium-niobate crystals (MgO:LiNbO3), as well as the optical local oscillators for homodyne detection. The green light field was mode-matched into the squeezed light resonators to pump degenerate type\,I parametric-down-conversion processes, which provided resonator output modes at 1064\,nm in squeezed vacuum states. The squeezed light resonators were singly-resonant for 1064\,nm, had a standing-wave and half-monolithic design, and a length of about 40\,mm. The resonators' coupling mirrors were attached to piezo electric transducers to stabilize the cavity lengths on resonance using the Pound-Drever-Hall locking scheme \cite{Drever83,Black}. An active temperature control stabilized the crystal temperatures at phase matching of the fundamental and the harmonic fields at about 60$^\circ$C.

\textit{Light detection and data aquisition:} The two-mode (squeezed) state was detected with two eight-port homodyne detectors incorporating altogether four conventional balanced homodyne detectors (BHDs). Each BHD consisted of a balanced beam splitter, two high-quantum efficiency PIN photo diodes and used a homodyne local oscillator power of about 3\,mW at 1064\,nm. The difference photo-electric current of the two photodiodes was amplified and transferred to a voltage by a trans-impedance amplifier. The voltage signal was then mixed with a 6.4\,MHz electronic local oscillator, anti-alias filtered with a corner frequency of 400\,kHz, and finally synchronously sampled at a frequency of 1\,MHz.

\textbf{Data availability statement:} The data that support these findings are available from the corresponding author upon request.

\subsection*{Acknowledgements}
This research was partially supported by the Deutsche Forschungsgemeinschaft (project SCHN 757/5-1).
J.F. acknowledges financial support from the EU FP7 under Grant Agreement
No. 308803 (project BRISQ2) cofinanced by MSMT CR (7E13032).
N.J.C. acknowledges financial support from the Fonds de la Recherche Scientifique (F.R.S.-FNRS) under grant T.0199.13 and from the Belgian Science Policy Office (BELSPO) under grant IAP P7-35 Photonics@be.
D.A. acknowledges financial support from the DFG Graduate School "Quantum mechanical noise in complex systems" (RTG1991) and thanks Tobias Gehring and Aiko Samblowski for helpful discussions. 

\subsection*{Author Contributions}
J.F.\ and N.J.C.\ developed the theory, D.A.\ and M.S.\ conducted the experiment and performed all measurements under the supervision of R.S. The data was analysed by J.F.\ and D.A.\ and J.F., D.A., R.S. and N.J.C. wrote the paper.

\subsection*{Competing financial interests}
The authors declare no competing financial interests.

\end{document}